\documentclass[
superscriptaddress,
preprintnumbers,
amsmath,amssymb,
aps,
twocolumn,
prb,
floatfix]{revtex4}
\usepackage{graphicx}
\usepackage{amssymb,amsmath,bbm,braket,indentfirst,bm}
\usepackage{dcolumn}
\allowdisplaybreaks
\usepackage{color}
\usepackage{bigstrut}

\newcommand{\e}[1]{\ensuremath \times 10^{#1}}

\newcommand{\pder}[2]{\ensuremath \frac{\partial#1}{\partial#2}} 

\renewcommand\({\ensuremath \left(}
\renewcommand\){\ensuremath \right)}
\renewcommand\[{\ensuremath \left[}
\renewcommand\]{\ensuremath \right]}

\preprint{MIT/CTP-4473}
\begin{document}

\title{Generalized energy and time-translation invariance in a driven, dissipative system}

\author{Thomas~Iadecola}
\affiliation{Physics Department, Boston University, Boston, Massachusetts 02215, USA}

\author{Claudio~Chamon}
\affiliation{Physics Department, Boston University, Boston, Massachusetts 02215, USA}

\author{Roman~Jackiw}
\affiliation{Department of Physics, Massachusetts Institute of Technology, Cambridge, Massachusetts 02139, USA}

\author{So-Young~Pi}
\affiliation{Physics Department, Boston University, Boston, Massachusetts 02215, USA}

\date{\today}

\begin{abstract}
Driven condensed matter systems consistently pose substantial challenges to theoretical understanding.  Progress in the study of such systems has been achieved using the Floquet formalism, but certain aspects of this approach are not well understood.  In this paper, we consider the exceptionally simple case of the rotating Kekul\'e mass in graphene through the lens of Floquet theory.  We show that the fact that this problem is gauge-equivalent to a time-independent problem implies that the ``quasi-energies" of Floquet theory correspond to a continuous symmetry of the full time-dependent Lagrangian.  We use the conserved Noether charge associated with this symmetry to recover notions of equilibrium statistical mechanics.
\end{abstract}

\maketitle

Driven quantum condensed matter systems have become a subject of great interest in recent years.\cite{hanggi1,hanggi2,platero}  Solid state systems in particular have attracted much attention due to the possibility of using external driving to engineer novel properties in materials.  Floquet theory,\cite{shirley,sambe,oka,kitagawa,fertig} which is one of the prevailing theoretical tools for studying such systems, has been used to achieve progress in this direction.  For instance, a class of materials known as ``Floquet topological insulators," which are normal materials that acquire topologically nontrivial features due to optical driving, has been proposed,\cite{lindner} and there is experimental evidence \cite{ftiexp} of its realization in photonic crystals.

Floquet theory is based on the following theorem, which is the time-domain analog of Bloch's theorem: if a Hamiltonian $H(t)$ is periodic in time, $H(t)=H(t+T)$ where $T$ is the period, then the solutions $\ket{\Psi_\alpha(t)}$ of the time-dependent Schr\"odinger equation $\[H(t)-i\partial_t\]\ket{\Psi_\alpha(t)}=0$ can be written as $\ket{\Psi_\alpha(t)}=e^{-i\epsilon_\alpha t}\ket{\Phi_\alpha(t)}$, where the Floquet states $\ket{\Phi_\alpha(t)}$ are also periodic in time, $\ket{\Phi_\alpha(t)}=\ket{\Phi_\alpha(t+T)}$.  The quantities $\epsilon_\alpha$, known as quasi-energies, are analogous to the crystal momenta of Bloch's theorem in that they are only well defined modulo the characteristic frequency $\Omega= 2\pi/T$.  It is also possible to define a time-independent Floquet effective Hamiltonian $H_{\rm eff}$ \cite{kitagawa,kitagawa2010,luca,jiang} whose eigenvalues are the quasi-energies $\epsilon_\alpha$ and which therefore inherits the multivaluedness of the quasi-energy spectrum.  Nevertheless, it is common practice to speak of the quasi-energy band structure \cite{lindner, kitagawa2010, rudner} of a system, which can be obtained by solving the Floquet eigenvalue problem $H_F\ket{\Phi_\alpha(t)}=\epsilon_\alpha\ket{\Phi_\alpha(t)}$, where $H_F\equiv H(t)-i\partial_t$ is known as the Floquet operator.

The apparent simplicity of Floquet theory belies certain conceptual difficulties.\cite{kohn}  In particular, because quasi-energies are only well defined modulo $\Omega$, there is no way of defining a lowest quasi-energy.  Consequently, there is no notion of the ground state of a driven system in Floquet theory.  For systems placed in contact with a heat bath at finite temperature, it is possible to derive master equations for the time evolution of the reduced density matrix.\cite{blumel,hanggi3,hanggi4}  This approach is only practicable on a case-by-case basis and frequently involves the use of various approximations, so that no general and intuitive notion of the occupation of a Floquet state exists for such systems.  The inclusion of dissipation for many-body systems remains quite challenging.

In this paper, we study an exactly solvable model where these questions have clear answers.  We consider the steady state reached by a system of Dirac fermions in graphene when coupled to a heat bath of acoustic phonons in the presence of a rotating Kekul\'e mass term.    This problem can be solved exactly both with and without Floquet theory.  Without Floquet theory, one can solve it by a mapping to a time-independent system via an axial gauge transformation, which preserves all transport properties of the system.\cite{rotatingkekule}  In this work, we illustrate the equivalence of this approach to that of Floquet theory.  In particular, we show that the quasi-energies defined above correspond in a simple way to the energy eigenvalues of the time-independent Hamiltonian $\tilde{\mathcal H}$ of Eq.\ \eqref{htilde}, which describes the gauge-transformed system.  In this way, we see that this time-independent Hamiltonian is indeed a Floquet effective Hamiltonian $H_{\rm eff}$ in the sense of Refs.\ \onlinecite{kitagawa} and \onlinecite{kitagawa2010,luca,jiang}. We subsequently illustrate that these quasi-energies correspond to a conserved Noether charge in the time-dependent problem.  This Noether charge differs from what we normally call ``energy" in that it is associated with a generalized time-translation symmetry, which involves both a shift in time and a compensating chiral rotation of the Dirac spinors.  We also show that it is possible to use this Noether charge to recover notions of equilibrium statistical mechanics by constructing an ensemble governing the probability distribution of the various states accessible to the system when the bath is held at finite temperature.  At zero temperature, the ``ground state" of the system can be determined by minimizing this generalized energy.  


We consider a time-dependent Lagrangian of the form $\mathcal L_{\rm tot}=\mathcal L_{\rm sys}+\mathcal L_{A_5}+\mathcal L_{\rm bath}$, where
\begin{subequations}\label{tlagrangian}
\begin{align}
\mathcal L_{\rm sys}&= \bar\Psi\[\gamma^\mu\; i\partial_\mu - |\Delta|\; e^{-i\gamma_5(\Omega t+\varphi)}\]\Psi \label{lsys}\\
\mathcal L_{A_5}&= j_5^i\; A_{5\; i}=j_5^i\; \bar A_{5\; i}+j_5^i\; \delta A_{5\; i}\nonumber\\
&=\mathcal L_{\rm strain}+\mathcal L_{\rm sys-bath}\label{la}\\
\mathcal L_{\rm bath}&= \frac{M}{2}|\dot{\bm u}|^2-\frac{1}{2}\; C_{ijkl}\;  u_{ij}\;  u_{kl}\label{lbath}\ .
\end{align}
\end{subequations}
The physics of each term is briefly explained below.  $\mathcal L_{\rm sys}$ is the low-energy Dirac field theory for fermions hopping on a hexagonal lattice against the background of a particular phonon mode with wave vector $\bm K_+$ and energy $\Omega$.\cite{rotatingkekule}  This phonon mode leads to the time-dependent mass term in \eqref{lsys}, whose magnitude is controlled by the magnitude of the complex Kekul\'e order parameter $\Delta=|\Delta|e^{i\varphi}$.  We use Dirac spinors $\Psi^\dagger_{\bm p} = (b_{+,\bm p}^\dagger\; a_{+,\bm p}^\dagger\; a_{-,\bm p}^\dagger\; b_{-,\bm p}^\dagger)$, where $a_{\pm,\bm p}^\dagger$ creates a fermion on sublattice $A$ with momentum $\bm p$ and the chiral indices $\pm$ label the valley (and similarly for $b^\dagger_{\pm,\bm p}$).  Our Dirac matrices are
\begin{align*}
\gamma^0=\begin{pmatrix}0&\mathbbm1\\\mathbbm1&0\end{pmatrix}\ ,\indent \gamma^i = \begin{pmatrix}0&-\sigma_i\\ \sigma_i&0\end{pmatrix}\ ,\indent \gamma_5 = \begin{pmatrix}\mathbbm 1&0\\0&-\mathbbm 1\end{pmatrix}\ ,
\end{align*}
where $\mathbbm 1$ is the $2\times 2$ identity matrix and $\sigma_i$, $i=1,2$ are Pauli matrices, and we use the standard notation $\bar{\Psi}\equiv\Psi^\dagger\gamma^0$.  The fermions in \eqref{lsys} couple to the spatial components \footnote{We take $A_{5\; 0}=0$ without loss of generality.} of an axial gauge field $A_{5\; \mu}$ in Eq.\ \eqref{la} through the axial current operator $j_5^i\equiv \bar\Psi\gamma^i\gamma_5\Psi$.  By calculating the changes in nearest-neighbor hoppings due to uniaxial strain (see Refs.\ \onlinecite{rotatingkekule} and \onlinecite{solitons}) and linearizing the resulting Hamiltonian around the Dirac points, it can be shown that the fields $A_{5\; i}$ depend explicitly on the fields $u_i(\bm x,t)$ in Eq.\ \eqref{lbath}, which measure locally the average displacement of the lattice sites from their equilibrium positions, via the relations
\begin{align*}
A_{5\; 1}=\frac{\alpha t_0}{d}\; \frac{3}{2}\( u_{22}- u_{11}\)\ , \indent A_{5\; 2}=\frac{\alpha t_0}{d}\; \frac{3}{2}\( u_{12}+ u_{21}\)\ ,
\end{align*}
where the strain field $ u_{ij}(\bm x,t)\equiv \(\partial_iu_j+\partial_ju_i\)/2$.  Here, $\alpha\approx 3.7$ is the dimensionless electron-phonon coupling, $t_0\approx 2.8\text{ eV}$ is the uniform hopping amplitude in the absence of strain, and $d\approx 1.4\text{ \AA}$ is the nearest-neighbor spacing.  The $A_{5\; i}$ thus encode the effects of strain in the graphene lattice---under constant uniaxial strain, they acquire a constant value $\bar A_{5\; i}$, around which there exist small time- and space- dependent fluctuations $\delta A_{5\; i}(\bm x,t)$ due to acoustic phonons.  We take these acoustic phonons to constitute a heat bath which allows the fermions in \eqref{lsys} to achieve a steady state in the presence of the time-dependent Kekul\' e mass term, so that $\mathcal L_{\rm sys-bath}$ in \eqref{la} constitutes a system-bath interaction.  The Lagrangian for the bath is written in terms of the $u_i(\bm x,t)$ in Eq.\ \eqref{lbath}, where in the first term $M$ sets the kinetic energy scale of the acoustic phonons and in the second term the elastic tensor $C_{ijkl}$ encodes the energy cost of strain along different directions.

The single-particle Hamiltonian corresponding to \eqref{lsys} is given by
\begin{align}\label{hsys}
\mathcal H_{\rm sys}(t)&=\begin{pmatrix}\bm \sigma\cdot\bm p &\Delta\; e^{i\Omega t}\; \mathbbm 1\\ \Delta^*\; e^{-i\Omega t} \; \mathbbm 1&-\bm \sigma\cdot\bm p \end{pmatrix},
\end{align}
where $\bm p=(p_x,p_y)$ is the momentum operator, and where the $2\times 2$ identity matrix $\mathbbm 1$ and the Pauli matrices $\bm\sigma=(\sigma_1,\sigma_2)$ act on sublattice indices.

Before proceeding, it is interesting to note that the Hamiltonian \eqref{hsys} bears a striking resemblance to the famous Rabi problem,\cite{rabi} which concerns a single spin in a magnetic field that rotates about the $z$-axis.  Here, the order parameter $\Delta$ plays the role of the magnetic field, and the valley degree of freedom plays the role of spin.  Note, however, that the problem described by the Hamiltonian \eqref{hsys} differs from the Rabi problem in several important ways.  For example, \eqref{hsys} describes a system of many non-interacting fermions, rather than a two-state system.  Furthermore, the dispersion in the kinetic term implies that the resonance condition varies with $p\equiv|\bm p|$, so that the system has no single resonant frequency.

The time-dependent problem governed by Eq.s \eqref{tlagrangian} is particularly simple in that all explicit time dependence can be removed by defining
\begin{align}\label{gtrans}
\tilde\Psi = e^{-i\gamma_5\frac{\Omega t}{2}}\Psi,&\hspace{.25cm} \tilde A_{5\; 0}=-\frac{\Omega}{2},\hspace{.25cm} \tilde A_{5\; i}=A_{5\; i},\nonumber\\
\tilde u_i&=u_i,\hspace{.3 cm} \tilde u_{ij}= u_{ij}.
\end{align}
This amounts to a time-dependent axial gauge transformation of $\mathcal L_{\rm tot}$ that maps the problem into a ``rotating frame."  The transformed Lagrangian for the fermions is found to be
\begin{align}\label{ltilde}
\tilde{\mathcal L}_{\rm sys} = \bar{\tilde{\Psi}}\(\gamma^\mu\; i\partial_\mu -\gamma^0\gamma_5\; \frac{\Omega}{2}-|\Delta|\; e^{-i\gamma_5\varphi}\)\tilde\Psi\ ,
\end{align}  
with the corresponding single-particle Hamiltonian given in matrix form by
\begin{align}\label{htilde}
\tilde{\mathcal H} = \begin{pmatrix}\bm \sigma\cdot\bm p +\frac{\Omega}{2}\; \mathbbm 1&\Delta\; \mathbbm 1\\ \Delta^*\; \mathbbm 1&-\bm \sigma\cdot\bm p - \frac{\Omega}{2}\; \mathbbm 1\end{pmatrix}.
\end{align}
The transformation \eqref{gtrans} ensures that $\tilde{\mathcal L}_{A_5}=\mathcal L_{A_5}$, and $\tilde{\mathcal L}_{\rm bath}=\mathcal L_{\rm bath}$.  Moreover, \eqref{gtrans} has no effect on the bath or the system-bath coupling. Therefore, the eigenstates of the Hamiltonian \eqref{htilde} are thermal states described by a density operator $\tilde\rho = \exp(-\beta\tilde{\mathcal H})/\text{tr }[\exp(-\beta\tilde{\mathcal H})]$, where $1/\beta$ is the temperature of the bath.  Since the U(1) current operator $j^\mu = \bar\Psi\gamma^\mu\Psi$ is also invariant under \eqref{gtrans}, all transport properties of $\mathcal L_{\rm sys}$ are identical to those of the Hamiltonian \eqref{htilde}.  Because the spectrum of $\tilde{\mathcal H}$, which is given by the four energy bands $E_{\pm,\pm}=\pm\sqrt{(p\pm\Omega/2)^2+|\Delta|^2}$, is that of a semiconductor with gap $2|\Delta|$, we conclude that the same must be true of the time-dependent system described by $\mathcal L_{\rm sys}$.  Taken together, these observations show that the transformation \eqref{gtrans} maps the non-equilibrium steady state of $\mathcal L_{\rm tot}$ to a thermal state in the rotating frame, and that the transport properties of this thermal state are identical to those of the full time-dependent problem.

The rotating Kekul\'e mass problem is also strikingly simple from the point of view of Floquet theory, as we now show.  We wish to solve the Floquet eigenvalue problem $\mathcal H_F\ket{\Phi_\alpha(t)}=\epsilon_\alpha\ket{\Phi_\alpha(t)}$, where $\mathcal H_F =\mathcal H_{\rm sys}-i\partial_t$ and $\mathcal H_{\rm sys}$ is the Hamiltonian corresponding to $\mathcal L_{\rm sys}$, defined in \eqref{lsys}.  To do this, we note that because the Floquet states $\ket{\Phi_\alpha(t)}$ are periodic with frequency $\Omega$, we may expand them in a Fourier series: $\ket{\Phi_\alpha(t)}=\sum_{n=-\infty}^\infty e^{-in\Omega t}\ket{\Phi_\alpha^n}$.  Substituting this into the Floquet equation and applying the operator $\frac{1}{T}\int_0^Tdt\; e^{im\Omega t}$ to both sides, we obtain the equation
\begin{align}\label{floquet}
\sum_{n=-\infty}^\infty \(\mathcal H_{mn}-n\Omega\; \delta_{mn}\)\ket{\Phi^n_\alpha}=\epsilon_\alpha\ket{\Phi^m_\alpha}\ ,
\end{align}
where $\mathcal H_{mn}=\frac{1}{T}\int_0^Tdt\; e^{i(m-n)\Omega t}\; \mathcal H_{\rm sys}(t)$.  Because $\mathcal H_{\rm sys}$ contains only one harmonic of the driving frequency $\Omega$, we have
\begin{subequations}\label{h}
\begin{align}
\mathcal H_{m m}&\equiv \mathcal H_0 =\begin{pmatrix}\bm \sigma\cdot\bm p&0\\0&-\bm \sigma\cdot\bm p\end{pmatrix}\\
\mathcal H_{m\, m+1}&\equiv \mathcal H_1=\begin{pmatrix}0&\Delta\;\mathbbm 1\\0&0\end{pmatrix}\\
\mathcal H_{m\, m-1}&\equiv \mathcal H_{-1}=\begin{pmatrix}0&0\\\Delta^*\;\mathbbm 1&0\end{pmatrix}\\
\mathcal H_{mn}&=0\indent\text{if $|m-n|>1$.}
\end{align}
\end{subequations}
\vskip-.1cm
\noindent In matrix form \eqref{floquet} becomes
\begin{widetext}
\begin{align}\label{floquetmatrix}
\begin{pmatrix}\ddots&\vdots&\vdots&\vdots& \\
\cdots & \mathcal H_0-\Omega\; \mathbbm 1&\mathcal H_1&0&\cdots&\\
\cdots & \mathcal H_{-1}&\mathcal H_0&\mathcal H_1&\cdots\\
\cdots&0&\mathcal H_{-1}&\mathcal H_0+\Omega\; \mathbbm 1&\cdots\\
&\vdots&\vdots&\vdots&\ddots\end{pmatrix}\begin{pmatrix}\vdots\\ \ket{\Phi_\alpha^1}\\ \ket{\Phi^0_\alpha}\\ \ket{\Phi^{-1}_\alpha}\\ \vdots\end{pmatrix} = \epsilon_\alpha\begin{pmatrix}\vdots\\ \ket{\Phi_\alpha^1}\\ \ket{\Phi^0_\alpha}\\ \ket{\Phi^{-1}_\alpha}\\ \vdots\end{pmatrix}\ ,
\end{align}
\end{widetext}
where now $\mathbbm 1$ is a $4\times 4$ identity matrix.  Eq.\ \eqref{floquetmatrix} defines an eigenvalue problem for an infinite-dimensional matrix, which is impossible to solve in general.  The standard approach from this point forward is to truncate the number of harmonics at some $m=\pm m_0$ and let the sum in \eqref{floquet} run from $-m_0$ to $m_0$ in order to find the quasi-energy spectrum in the truncated space (see, e.g., Refs.\ \onlinecite{oka}, \onlinecite{kitagawa} and \onlinecite{lindner}).  In fact, for the Floquet matrix \eqref{floquetmatrix}, one can prove by construction (see Appendix) that for a given $m_0>0$ the $4(2m_0+1)$ quasi-energy eigenvalues are
\begin{subequations}\label{qespec}
\begin{align}
\epsilon^0_{\pm,\pm}&=\pm p\pm m_0\Omega \label{spurious}\\
\epsilon^n_{\pm,\pm,\pm}&=\pm\sqrt{(p\pm\Omega/2)^2+|\Delta|^2}\pm\frac{n\Omega}{2} \label{qes},
\end{align}
\end{subequations}
where $n=1,3,\dots,2m_0-1$.  Note that the degeneracy (mod $\Omega$) of the linearly-dispersing modes in \eqref{spurious} does not grow with $m_0$, while the degeneracy of the modes \eqref{qes} does.  This suggests that the quasi-energy modes listed in \eqref{spurious} are spurious artifacts of the truncation.  Indeed, one can show that the characteristic equation of the infinite Floquet matrix, namely $\det(\mathcal H_F-\lambda\; \mathbbm 1)=0$, is unchanged under the substitution $\lambda\to \lambda+n\Omega$, where $n$ is an integer. \cite{shirley,sambe}  From this, one concludes that if $\lambda$ is an eigenvalue of $\mathcal H_F$, then so is $\lambda+n\Omega$.  The eigenvalues in \eqref{qes} exhibit this periodicity mod $\Omega$, whereas the eigenvalues in \eqref{spurious} do not.  We conclude that the latter modes are indeed spurious, and we take \eqref{qes}, with $n$ any positive odd integer, to constitute the true quasi-energy spectrum of $\mathcal H_F$.  The spectrum of the system's Floquet effective Hamiltonian $\mathcal H_{\rm eff}$ is obtained by choosing a single quasi-energy branch, say $n=1$ without loss of generality.  Once this choice of branch is made, however, we see that the quasi energies $\epsilon_{\pm,\pm,+}^1$ are identical to the energy eigenvalues $E_{\pm,\pm}$ of $\tilde{\mathcal H}$, up to a constant shift by $-\Omega/2$ which is unimportant.  This indicates that the rotating-frame Hamiltonian $\tilde{\mathcal H}$ of Eq.\ \eqref{htilde} can in fact be identified with a Floquet effective Hamiltonian of the system.

The time-dependent Hamiltonian $\mathcal H_{\rm sys}$ of Eq.\ \eqref{hsys} is related to the time-independent Floquet effective Hamiltonian $\tilde{\mathcal H}$ of Eq.\ \eqref{htilde} by a unitary transformation
\begin{align*}
\tilde{\mathcal H}=U(t)\mathcal H_{\rm sys}(t)U^\dagger(t)-iU(t)\partial_tU^\dagger(t)\ ,
\end{align*} 
where $U(t)=e^{-i\gamma_5\Omega t/2}$.  This implies that its quasi-energy spectrum, which we now identify with the energy spectrum $E_{\pm,\pm}$ of $\tilde{\mathcal H}$, corresponds to a continuous symmetry of the Lagrangian $\mathcal L_{\rm sys}$ according to the following argument.  The Lagrangian $\tilde{\mathcal L}_{\rm sys}$ of Eq.\ \eqref{ltilde}, which corresponds to the rotating-frame Hamiltonian $\tilde{\mathcal H}$, has no explicit time dependence.  Under an infinitesimal time translation $t\to t-a$, we have $\delta\tilde\Psi=\partial_t\tilde\Psi$, so that to leading order in $\delta\tilde\Psi$ one obtains $\delta\tilde{\mathcal L}_{\rm sys}=\partial_t\tilde{\mathcal L}_{\rm sys}$, indicating that $t\to t-a$ is (expectedly) a symmetry of the action associated with $\tilde{\mathcal L}_{\rm sys}$.  A standard calculation using Noether's theorem shows that the conserved quantity associated with this symmetry is indeed the Hamiltonian $\tilde{\mathcal H}$.  However, we can also use the relation $\tilde{\Psi} = e^{-i\gamma_5\Omega t/2}\; \Psi$ from Eq.\ \eqref{gtrans} to obtain a corresponding transformation law for $\Psi$, namely
$\delta\Psi = \partial_t\Psi -i\gamma_5(\Omega/2)\Psi$.  This is the variation in $\Psi$ brought about by the infinitesimal version of the continuous symmetry
\begin{align}\label{symm}
t\to t-a,\indent \Psi(t)\to e^{-i\gamma_5\Omega a/2}\; \Psi(t+a),
\end{align}
which combines a time translation with a compensating chiral rotation of the Dirac spinors.  Because $\mathcal L_{A_5}$ and $\mathcal L_{\rm bath}$ are invariant under chiral rotations of the spinors, we see that \eqref{symm} is simply a time translation from the point of view of these terms in $\mathcal L_{\rm tot}$.  Consequently, the remaining fields in $\mathcal L_{\rm tot}$ transform as $\delta A_{5\; i}=\partial_tA_{5\; i}$, $\delta u_i=\partial_t u_i$, and $\delta u_{ij}=\partial_t u_{ij}$ under \eqref{symm}.\footnote{Note that because $\tilde{\mathcal L}_{A_5}=\mathcal L_{A_5}$ and $\tilde{\mathcal L}_{\rm bath}=\mathcal L_{\rm bath}$, these transformation rules also hold in the rotating frame.}  It is easily verified using these definitions, along with that of $\delta\Psi$ above, that $\delta\mathcal L_{\rm tot}=\partial_t\mathcal L_{\rm tot}$ under \eqref{symm}.  We conclude that \eqref{symm} is indeed a symmetry of the action associated with $\mathcal L_{\rm tot}$.  A straightforward calculation shows that the Noether charge corresponding to this symmetry is
\vskip -1cm
\begin{widetext}
\begin{align}\label{qtot}
Q_{\rm tot} &=\int d^2x\; \left\{\Psi^\dagger\[\alpha^ip_i+\gamma^0|\Delta|\; e^{-i\gamma_5(\Omega t+\varphi)}+\gamma_5\; \frac{\Omega}{2}\]\Psi-j_5^iA_{5\; i}+\frac{M}{2}|\dot{\bm u}|^2+\frac{1}{2}C_{ijkl}\;  u_{ij} u_{kl}\right\}\ ,
\end{align}
\end{widetext}
where $\alpha^i\equiv \gamma^0\gamma^i$.  It can be checked using the Euler-Lagrange equations of motion that $\partial_t Q_{\rm tot}=0$.  We partition $Q_{\rm tot}$ into contributions from the system, the bath, and the system-bath coupling as follows:
\begin{subequations}
\begin{align}
Q_{\rm sys}&= \int d^2x\; \Bigl\{\Psi^\dagger\Bigl[\alpha^ip_i+\gamma^0|\Delta|\; e^{-i\gamma_5(\Omega t+\varphi)} \nonumber\\
&\indent  +\gamma_5\; \frac{\Omega}{2}\;\Bigr]\Psi-j_5^i\bar A_{5\; i} \Bigr\}\label{qsys}\\
Q_{\rm sys-bath}&=-\int d^2x\; j_5^i\; \delta A_{5\, i}\label{qsysbath}\\
Q_{\rm bath}&= \int d^2x\; \(\frac{M}{2}|\dot{\bm u}|^2+\frac{1}{2}C_{ijkl}\;  u_{ij} u_{kl}\).\label{qbath}
\end{align}
\end{subequations}
$Q_{\rm sys}$, which contains contributions from $\mathcal L_{\rm sys}$ and $\mathcal L_{\rm strain}$, defined in \eqref{lsys} and \eqref{la}, is a generalized energy corresponding to the generalized time-translation symmetry \eqref{symm}.  Because this transformation is just a time translation from the point of view of the bath and the system-bath interaction, the quantities $Q_{\rm bath}$ and $Q_{\rm sys-bath}$ are identical to the physical energies corresponding to the familiar time-translation invariance.

The foregoing arguments have yielded a conserved quantity involving system and bath degrees of freedom.  The fact that we have such a conserved quantity suggests that we can construct an equilibrium statistical ensemble for calculating thermodynamic averages.  We now make these ideas more precise.  We begin by noting that in the absence of a system-bath coupling, the quantity $Q_{\rm sys}$ of Eq.\ \eqref{qsys} is conserved.  When the system is coupled to the bath, the conserved quantity is $Q_{\rm tot}=Q_{\rm sys}+Q_{\rm sys-bath}+Q_{\rm bath}$.  Because the fluctuations in $A_{5\; i}$ due to the acoustic phonons are small, we may approximate $Q_{\rm tot}\approx Q_{\rm sys}+Q_{\rm bath}$.  In this picture, the system's ``energy" $Q_{\rm sys}$ is no longer conserved---instead, the system and bath exchange this ``energy", while $Q_{\rm tot}$ remains constant.  Suppose we wish to determine the probability $P(Q_n)$ of the system having $Q_{\rm sys}=Q_n$, where $n$ labels the state of the system, for a fixed $Q_{\rm tot}$.  If we assume, in the spirit of equilibrium statistical mechanics, that this probability is proportional to the number $\mathcal N(Q_{\rm bath})=\mathcal N(Q_{\rm tot}-Q_n)$ of states available to the bath for a fixed value of $Q_{\rm bath}$, then we can deduce the form of $P(Q_n)$ as follows.  If we assume that $Q_n\ll Q_{\rm tot}$, then we can expand $\ln \mathcal N(Q_{\rm bath})$ around $Q_{\rm bath}=Q_{\rm tot}$ to obtain
\vskip -1.1cm
\begin{widetext}
\begin{align}
\ln\mathcal N(Q_{\rm bath})&= \ln \mathcal N(Q_{\rm tot})+\pder{\ln\mathcal N(Q_{\rm bath})}{Q_{\rm bath}}\Bigg\vert_{Q_{\rm bath}=Q_{\rm tot}}\hspace{-.5cm}(Q_{\rm bath}-Q_{\rm tot})+\ldots\nonumber\\
&=\ln \mathcal N(Q_{\rm tot})-\pder{\ln\mathcal N(Q_{\rm bath})}{Q_{\rm bath}}\Bigg\vert_{Q_{\rm bath}=Q_{\rm tot}}\hspace{-.5cm}Q_n+\mathcal O(Q_n^2)\label{entropy}
\end{align}
\end{widetext}
Recalling our earlier observation that $Q_{\rm bath}$ is just the energy of the bath, we see that the coefficient of $Q_{n}$ in \eqref{entropy} can be identified with the familiar Lagrange multiplier $\beta$, where $1/\beta$ is the temperature of the bath. We conclude that $P(Q_n)$ assumes the form of a Boltzmann distribution with the usual system energies $E_n$ replaced by the generalized energies $Q_n$:
\begin{align}
P(Q_n)=\frac{e^{-\beta Q_n}}{\sum_{n}e^{-\beta Q_n}},
\end{align}
where $n$ runs over all states accessible to the system.  

In summary, we have presented in this paper a study of the rotating Kekul\'e mass in graphene from the point of view of Floquet theory.  We found that the time-independent Hamiltonian \eqref{htilde} of the system in the rotating frame is in fact a Floquet effective Hamiltonian whose eigenvalues are quasi-energies.  Exploiting the fact that the time-dependent Hamiltonian of the system is related to this effective Hamiltonian by a unitary transformation, we showed that these quasi-energies correspond to a continuous symmetry \eqref{symm}, with an associated Noether charge $Q$, of the time-dependent Hamiltonian.  By explicitly coupling the system to a heat bath consisting of acoustic phonons, we constructed a statistical ensemble governing the probabilities of the various microstates accessible to the system.  In this way, we recover notions of equilibrium statistical mechanics, despite the fact that the original problem of Eq.\ \eqref{tlagrangian} is manifestly out of equilibrium.    In principle, such a construction is possible for any time-dependent Hamiltonian that is related by a unitary transformation to a time-independent Hamiltonian.  Further investigations of this exceptional class of problems and its generalizations could be enormously beneficial to the study of non-equilibrium quantum systems.

We thank Luca D'Alessio, Herb Fertig, and Takashi Oka for helpful discussions.  This work is
supported by DOE grants DEF-06ER46316 (T.I. and C.C.), DE-FG02-05ER41360 (R.J.), and DE-SC0010025 (S-Y. P.).

\appendix*
\begin{widetext}
\section{}

We present here a constructive proof of the assertion that truncating the Floquet matrix $\mathcal H_F$ at some $m=\pm m_0$ yields the quasi-energy spectrum of Eq.s \eqref{qespec}.  The truncated eigenvalue problem reads $\mathcal H_F^{(m_0)}\Phi_\alpha^{(m_0)}=\epsilon_\alpha^{(m_0)}\Phi^{(m_0)}_\alpha$, where
\begin{align*}
\mathcal H_F^{(m_0)}=\begin{pmatrix}
\mathcal H_0-m_0\Omega\; \mathbbm 1&\mathcal H_1&0&\cdots\\
\vspace{-.18cm}\mathcal H_{-1}&\mathcal H_0-(m_0-1)\Omega\; \mathbbm 1&\mathcal H_1&\cdots\\
\vspace{-.18cm}0&\mathcal H_{-1}&\hspace{-1cm}\ddots&\hspace{-1.5cm}\ddots\\
\vdots&\hspace{1cm}\ddots&\hspace{.25cm}\ddots&\mathcal H_1\\
0&\hspace{-2cm}\cdots&\hspace{-1.25cm}\mathcal H_{-1}&\mathcal H_0+m_0\Omega\; \mathbbm 1
\end{pmatrix},
\end{align*}
with $\mathcal H_0$ and $\mathcal H_{\pm 1}$ given in Eq.s \eqref{h}.  We construct the $4(2m_0+1)$ eigenvectors of $\mathcal H_F^{(m_0)}$ in the following way.  We begin by noting that the $4\times 4$ matrix $\mathcal H_0$ has eigenvalues $\pm p$ (each with multiplicity 2) and unnormalized eigenvectors
\begin{align}\label{es}
e_\pm=\begin{pmatrix}0\\0\\\pm e^{-i\theta}\\1\end{pmatrix}\indent\text{and}\indent e^\prime_{\pm}=\begin{pmatrix}
\pm e^{-i\theta}\\
1\\
0\\
0
\end{pmatrix}\ .
\end{align}
These in turn are eigenvectors of the matrices $\mathcal H_0\pm m_0\Omega\;\mathbbm 1$ with eigenvalues $\pm p\pm m_0\Omega$.  Furthermore, because $\mathcal H_{-1}\; e_\pm=\mathcal H_1\; e^\prime_\pm=0$, one verifies that the block-form vectors
\begin{align}\label{spur}
\Phi_{\pm,-}^{0,(m_0)}=\begin{pmatrix}
e_\mp\\0_4\\\vdots\\0_4
\end{pmatrix}\indent\text{and}\indent\Phi_{\pm,+}^{0,(m_0)}=\begin{pmatrix}0_4\\\vdots\\0_4\\e^\prime_{\pm}\end{pmatrix},
\end{align}
where $0_4$ is a 4-dimensional column vector of zeroes, are eigenvectors of $\mathcal H_F^{(m_0)}$ with eigenvalues $\epsilon^0_{\pm,\pm}$ given by Eq. \eqref{spurious}.  The remaining $8m_0$ eigenvectors are constructed by considering the following generic $8\times8$ sub-block of $\mathcal H_F^{(m_0)}$:
\begin{align}\label{sub}
\mathcal H_{8}^{(n)}&=\begin{pmatrix}
\mathcal H_0-n\Omega\;\mathbbm1&\mathcal H_1\\
\mathcal H_{-1}&\mathcal H_0-(n-1)\Omega\;\mathbbm1
\end{pmatrix},
\end{align}
where $-(m_0-1)\leq n\leq m_0$.  One can show that the four relevant eigenvalues and eigenvectors of $\mathcal H_8^{(n)}$ are
\begin{align*}
\mathcal H_8^{(n)}\begin{pmatrix}f_{\pm,-}(p)\; e^\prime_+\\ e_+\end{pmatrix}&=\epsilon^{2n-1}_{\pm,-,-}\begin{pmatrix}f_{\pm,-}(p)\; e^\prime_+\\ e_+\end{pmatrix}\\
\mathcal H_8^{(n)}\begin{pmatrix}-f_{\mp,+}(p)\; e^\prime_-\\ e_-\end{pmatrix}&=\epsilon^{2n-1}_{\pm,+,-}\begin{pmatrix}-f_{\mp,+}(p)\; e^\prime_-\\ e_-\end{pmatrix}.
\end{align*}
Here,
\begin{align*}
f_{\pm,\mp}(p)&=\frac{1}{2\Delta^*}\[(2p\mp \Omega)\pm\sqrt{(2p\mp\Omega)^2+4|\Delta|^2}\],
\end{align*}
$e_{\pm}$ and $e^\prime_{\pm}$ are as in \eqref{es}, and $\epsilon^{2n-1}_{\pm,\pm,-}$ are defined as in Eq.\ \eqref{qes}.  Moreover, because as previously noted $\mathcal H_{-1}\; e_\pm=\mathcal H_1\; e^\prime_\pm=0$, one verifies that the block-form vectors
\begin{align}\label{realtop}
\Phi^{2m_0-1,(m_0)}_{\pm,-,-}&=\begin{pmatrix}f_{\pm,-}(p)\; e^{\prime}_+\\ e_+\\0_4\\\vdots\\0_4\end{pmatrix}\indent\text{and}\indent \Phi^{2m_0-1,(m_0)}_{\pm,+,-}=\begin{pmatrix}-f_{\mp,+}(p)\; e^{\prime}_-\\ e_-\\0_4\\\vdots\\0_4\end{pmatrix}
\end{align}
are eigenvectors of $\mathcal H_F^{(m_0)}$ with eigenvalues $\epsilon^{2m_0-1}_{\pm,\pm,-}$ in the notation of Eq.\ \eqref{qes}.  Applying a similar argument to each sub-block of the form \eqref{sub} shows that the remaining $4(2m_0-1)$ eigenvectors are obtained by shifting the entries of the eigenvectors \eqref{realtop} downward by four positions at a time.  For instance, the eigenvectors corresponding to the quasi-energies $\epsilon^{2m_0-3}_{\pm,\pm,-}$ are given by
\begin{align*}
\Phi^{2m_0-3,(m_0)}_{\pm,-,-}&=\begin{pmatrix}0_4\\f_{\pm,-}(p)\; e^{\prime}_+\\ e_+\\0_4\\\vdots\\0_4\end{pmatrix}\indent\text{and}\indent \Phi^{2m_0-3,(m_0)}_{\pm,+,-}=\begin{pmatrix}0_4\\-f_{\mp,+}(p)\; e^{\prime}_-\\ e_-\\0_4\\\vdots\\0_4\end{pmatrix}.
\end{align*}
This shifting process can be repeated a total of $2m_0-1$ times, until we reach
\begin{align*}
\Phi^{2m_0-1,(m_0)}_{\pm,-,+}&=\begin{pmatrix}0_4\\\vdots\\0_4\\f_{\pm,-}(p)\; e^{\prime}_+\\ e_+\end{pmatrix}\indent\text{and}\indent \Phi^{2m_0-1,(m_0)}_{\pm,+,+}=\begin{pmatrix}0_4\\\vdots\\0_4\\-f_{\mp,+}(p)\; e^{\prime}_-\\ e_-\end{pmatrix},
\end{align*}
which correspond to the quasi-energies $\epsilon_{\pm,\pm,+}^{2m_0-1}$.  As a check that we have indeed exhausted all possible eigenvalues and eigenvectors of $\mathcal H_F^{(m_0)}$, we count 4 eigenvectors \eqref{spur} $+$ 4 eigenvectors \eqref{realtop} $+\ 4(2m_0-1)$ shifted eigenvectors $=8m_0+4$ eigenvectors in total, as desired.
\end{widetext}

\bibliography{floquet_kekule_paper.bib}
\end{document}